\documentclass[11pt]{article}
\usepackage{amsmath, amssymb, graphics}

\newcommand{\mathsym}[1]{{}}

\begin{document}

\title{Invariants and solutions of Gurzadyan-Xue dark energy cosmological
models}
\author{H.G.Khachatryan \\
ICRA, University 'La Sapienza', Rome, Italy\\ 
and Yerevan Physics Institute, Yerevan, Armenia 
}
\maketitle

\begin{abstract}
We have derived the invariants of cosmological models with 
Gurzadyan-Xue dark energy, along with the solutions for any time dependent light speed
and gravitational constant. The correspondence of the invariants with the 
separatrices found
earlier for the GX-models in \cite{Ver06a} is shown, and hence the basis of
then detected hidden symmetry is now revealed. Solutions are derived both
for radiation and matter models, as well as with both components.
GX-invariants act as efficient tools describing the models and the phases of
the cosmological expansion.
\end{abstract}

The formula for dark energy derived by Gurzadyan and Xue predicts the
observed SN value without any free parameters 
\cite{GX}. That formula written for the cosmological constant is 
\begin{equation}
\Lambda_{GX} =\frac{\pi ^{2}c^{2}(t)}{a^{2}(t)}  \label{I.1}
\end{equation}%
where $c(t)$ is the time dependent speed of light, $a(t)$ a scale factor
(length scale of current Universe). The GX-vacuum energy density related to $%
\Lambda $ reads 
\begin{equation}
\mu _{\Lambda}=\frac{\Lambda }{8\pi G(t)}=\frac{\pi c^{2}(t)}{8G(t)a^{2}(t)}%
,  \label{I.2}
\end{equation}%
where $G(t)$ is time dependent gravitational constant. In General Relativity
all the quantities in eq.(\ref{I.1}) and eq.(\ref{I.2}) are constants,
except for the scale factor, which is a function of cosmic time. 
Adopting Gurzadyan-Xue (GX) scaling eq.(\ref{I.1}), one has to consider the
possibility of varying physical constants\footnote{
Generally speaking, the vacuum energy density is known to be constant only in
Minkowski spacetime.}.

Cosmological models based on this idea are considered in \cite{Ver06}. 
Then, in \cite{Ver06a} it was shown that the solutions of those models
have interesting behavior, in particular the presence of a separatrix in the
phase space of solutions was shown. That separatrix divides the solutions
into two classes: Friedmannian-like, i.e. with initial singularity and
non-Friedmannian solutions which begin with nonzero scale factor and
vanishing matter density. Each class of solutions is characterized by a
single quantity, the density parameter which is defined in the same way as
in the standard cosmological model 
\begin{equation}
\Omega _{m}=\frac{8\pi G_{0}}{3H_{0}^{2}}\mu _{0},  \label{I.3}
\end{equation}
where $\mu $ is matter density, $H$ is the Hubble constant, and index "$0$"
refers to the values today. It was shown that the separatrix in all models
is given by the density parameter $\Omega _{m}\approx 2/3$ (for k=0, i.e.
non-curved Universe), which indicated existence of some hidden simmetries
between the models, although cosmological equations look very different. In
this paper we discover the basis of that hidden symmetry. Namely, we have
derived the invariants, which, as is well known, has to posses the physical
system with a symmetry. The invariants are found for radiation dominated,
matter dominated cases, as well as for components.

{\bf Solutions for Gurzadyan-Xue cosmological models.}{\it Matter dominated Universe.}
For mass density and scale factor we have two equations \cite{Ver06b,Ver06c} 
\begin{eqnarray}
\dot{\mu}+3H(\mu +\frac{p}{c^{2}}) &=&-\dot{\mu}_{\Lambda }+(\mu +\mu
_{\Lambda })(\frac{2\dot{c}}{c}-\frac{\dot{G}}{G})  \nonumber \\
H^{2}+\frac{kc^{2}}{a^{2}}-\frac{\Lambda }{3} &=&\frac{8\pi G}{3}\mu
\label{MD.1}
\end{eqnarray}%
Here $\mu $ is matter mass density, $k=\pm 1,0$ is the spatial curvature.
For $p=0$ equation of state for matter, we have 
\begin{equation}
\dot{\mu}+3H\mu =-\dot{\mu}_{\Lambda }+(\mu +\mu _{\Lambda })(\frac{2\dot{c}%
}{c}-\frac{\dot{G}}{G})  \label{MD.2}
\end{equation}%
For matter mass density with GX-dark energy we have a solution 
\begin{equation}
\mu _{m}(t)=(b_{m}+\frac{\pi a(t)}{4})\frac{c^{2}(t)}{G(t)a^{3}(t)}
\label{MD.3}
\end{equation}%
Substituting it into eq.(\ref{MD.2}) and reminding that $H=\dot{a(t)}/a(t)$,
we have 
\begin{equation}
\dot{a}(t)=c(t)\sqrt{\frac{8\pi b_{m}}{3a(t)}+\pi ^{2}-k}  \label{MD.4}
\end{equation}%
In this representation, one can see that we have three type of models, i.e.
with $b_{m}>0$, $b_{m}<0$ and $b_{m}=0$.

First, consider the case when $b_{m}>0$. We get for the scale factor the
following equation 
\begin{eqnarray}
\sqrt{x(x+1)}+\ln (\sqrt{x+1}-\sqrt{x}) &=&\frac{3\sqrt{\pi ^{2}-k}}{8\pi
b_{m}}\int_{0}^{t}c(t^{\prime })dt^{\prime },  \nonumber \\
x &=&\frac{3(\pi ^{2}-k)a}{8\pi b_{m}}  \label{MD.5}
\end{eqnarray}%
To solve the eq.(\ref{MD.5}) one must define the dependence $c(t)$, the
speed of light as funcion of time.

Considering then the case $b=0$, we get for the matter density 
\begin{equation}
\mu _{m}(t)=\frac{\pi c^{2}(t)}{4G(t)a^{2}(t)}\equiv 2\mu _{\Lambda }(t),
\label{MD.6}
\end{equation}%
and for the scale factor 
\begin{equation}
a(t)=\sqrt{\pi ^{2}-k}\int_{0}^{t}c(t^{\prime })dt^{\prime }
\label{MD.7}
\end{equation}%
If we insert eq.(\ref{MD.7}) into (ref{I.1}), we have 
\begin{equation}
\Lambda =\frac{\pi ^{2}c^{2}}{a^{2}}=\frac{\pi ^{2}}{\pi ^{2}-k}(\frac{1}{%
c(t)}\int_{0}^{t}c(t^{^{\prime }})dt^{^{\prime }})^{-2}  \label{MD.Lambda}
\end{equation}

For the class of solutions with $b=0$ we have for $\Omega _{m}$ defined as 
\begin{equation}
\Omega _{m}=\frac{8\pi G_{0}\mu _{m0}}{3H_{0}^{2}}  \label{MD.8}
\end{equation}%
and for the current Hubble constant $H_{0}$ and 
\begin{equation}
H_{0}^{2}=\frac{c_{0}^{2}(\pi ^{2}-k)}{a_{0}^{2}}  \label{MD.9}
\end{equation}%
If we substitude eqs.(\ref{MD.9},\ref{MD.3}) into eq.(\ref{MD.8}), we have 
\begin{equation}
\Omega _{m}=\frac{2\pi ^{2}}{3(\pi ^{2}-k)}  \label{MD.10}
\end{equation}%
It's easy to see that for flat, $k=0$, model the last eq.(\ref{MD.10})
becomes 
\begin{equation}
\Omega _{m}=\frac{2}{3} ,  \label{MD.11}
\end{equation}%
i.e. the separatirix first discovered in \cite{Ver06a}. However, we now see
the reason of the hidden invariance discussed there.

Finally, for the case $b_{m}<0$, we have 
\begin{eqnarray}
\sqrt{x(x-1)}+\ln (\sqrt{x}+\sqrt{x-1}) &=&\frac{3\sqrt{\pi ^{2}-k}}{8\pi
|b_{m}|}\int_{0}^{t}c(t^{\prime })dt^{\prime },  \nonumber \\
x &=&\frac{3(\pi ^{2}-k)a}{8\pi |b_{m}|}  \label{MD.12}
\end{eqnarray}

{\it Radiation dominated Universe.}
For the equation of state, $3p/c^{2}=\mu $, we have 
\begin{equation}
\dot{\mu}_{r}+4H\mu _{r}=-\dot{\mu}_{\Lambda }+(\mu _{r}+\mu _{\Lambda })(%
\frac{2\dot{c}}{c}-\frac{\dot{G}}{G})  \label{MR.1}
\end{equation}%
For radiation mass density $\mu_r$ with GX-dark energy we obtain 
\begin{equation}
\mu _{r}(t)=(b_{r}+\frac{\pi a^{2}(t)}{8})\frac{c^{2}(t)}{G(t)a^{4}(t)}
\label{MR.2}
\end{equation}%
and from eq.(\ref{MR.1}) we have 
\begin{equation}
\dot{a}(t)=c(t)\sqrt{\frac{8\pi b_{r}}{3a^{2}(t)}+\frac{2\pi ^{2}}{3}-k}
\label{MR.3}
\end{equation}%
Consider again the three type of solutions, i.e. for $b_{r}>0$, $b_{r}<0$, $%
b_{r}=0$, respectively.

At $b_{r}>0$ we have 
\begin{eqnarray}
\sqrt{x^{2}+1} &=&\frac{(2\pi ^{2}-3k)}{\sqrt{24\pi b_{r}}}%
\int_{0}^{t}c(t^{\prime })dt^{\prime },  \nonumber \\
x &=&a\sqrt{\frac{(2\pi ^{2}-3k)}{8\pi b_{r}}}  \label{MR.4}
\end{eqnarray}%
And at $b_{r}<0$: 
\begin{eqnarray}
\sqrt{x^{2}-1} &=&\frac{(2\pi ^{2}-3k)}{\sqrt{24\pi |b_{r}|}}%
\int_{0}^{t}c(t^{\prime })dt^{\prime },  \nonumber \\
x &=&a\sqrt{\frac{(2\pi ^{2}-3k)}{8\pi |b_{r}|}}  \label{MR.5}
\end{eqnarray}%
At $b_{r}=0$ case, we get 
\begin{equation}
\mu _{r}(t)=\frac{\pi c^{2}(t)}{8G(t)a^{2}(t)}=\mu _{\Lambda }(t)
\label{MR.6}
\end{equation}%
For the scale factor of the latter group of models we have an equation 
\begin{equation}
a(t)=\sqrt{\frac{2}{3}\pi ^{2}-k}\int_{0}^{t}c(t^{^{\prime }})dt^{^{\prime }}
\label{MR.7}
\end{equation}%
and for cosmological constant 
\begin{equation}
\Lambda =\frac{3\pi ^{2}}{2\pi ^{2}-3k}(\frac{1}{c(t)}\int_{0}^{t}c(t^{^{%
\prime }})dt^{^{\prime }})^{-2}  \label{MR.8}
\end{equation}%
The separatrix value of $\Omega _{r}$ then is 
\begin{equation}
\Omega _{r}=\frac{\pi ^{2}}{(2\pi ^{2}-3k)}  \label{MR.9}
\end{equation}%
For flat models ($k=0$) we have 
\begin{equation}
\Omega _{r}=\frac{1}{2}  \label{MR.10}
\end{equation}

{\bf Gurzadyan-Xue invariants.}
Consider now for the mass density $\mu$(if the speed of light and
gravitational constant vary) in the equation 
\begin{equation}
\dot{\mu}+3H(\mu +\frac{p}{c^{2}})=-\dot{\mu}_{\Lambda }+(\mu +\mu _{\Lambda
})(\frac{2\dot{c}}{c}-\frac{\dot{G}}{G})  \label{1}
\end{equation}%
the form $\mu (t)=\mu _{r}(t)+\mu _{m}(t)$, where $\mu _{r}(t)$ and $\mu
_{m}(t)$ are defined as 
\begin{eqnarray}
\mu _{m}(t) &=&\alpha \mu (t),0\leq \alpha \leq 1  \nonumber \\
\mu _{r}(t) &=&(1-\alpha )\mu (t)  \label{2}
\end{eqnarray}%
$\mu _{r}(t)$ and $\mu _{m}(t)$ are radiation and matter mass densities,
respectively. For any smooth time-dependent function $f(t)$ we have 
\begin{equation}
\frac{df}{dt}=\frac{df}{da}\frac{da}{dt}=Ha\frac{df}{da},H=\frac{\dot{a}}{a}
\label{3}
\end{equation}%
If we take $p=0$ for the matter and $3p/c^{2}=\mu _{r}$ for the radiation,
after sustituding eq.(\ref{2}) into eq.(\ref{1}) in view of eq.(\ref{3}), we
have 
\begin{equation}
\acute{\mu}(a)+(4-\alpha )\mu (a)=-\acute{\mu}_{\Lambda }(a)+(\mu (a)+\mu
_{\Lambda }(a))(\frac{2\acute{c}(a)}{c(a)}-\frac{\acute{G}(a)}{G(a)})
\label{4}
\end{equation}%
After tedious derivations we obtain 
\begin{equation}
\frac{d}{da}(\ln \frac{\mu a^{4-\alpha }G}{c^{2}})=\frac{\mu _{\Lambda }}{%
\mu }\frac{d}{da}(\ln \frac{c^{2}}{G\mu _{_{\Lambda }}}),  \label{5}
\end{equation}%
in view of 
\begin{equation}
\Lambda =\frac{\pi ^{2}c^{2}}{a^{2}},\mu _{\Lambda }=\frac{\Lambda }{8\pi G}=%
\frac{\pi c^{2}}{8Ga^{2}}  \label{6}
\end{equation}%
Finally, we obtain the GX-invariant as 
\begin{equation}
\frac{\mu a^{4-\alpha }G}{c^{2}}-\frac{\pi a^{2-\alpha }}{4(2-\alpha )}%
=b_{mr}  \label{7}
\end{equation}%
If we demand $\mu $ to be maximal, i.e. 
\begin{eqnarray}
\frac{\partial \mu }{\partial \alpha } &=&0,  \nonumber \\
\frac{\partial ^{2}\mu }{\partial \alpha ^{2}} &>&0.  \label{8}
\end{eqnarray}%
we have 
\begin{eqnarray}
\frac{\pi a^{2-\alpha }}{4(2-\alpha )^{2}}+b_{mr}\ln (a) &=&0,  \nonumber \\
b_{mr} &>&0  \label{9}
\end{eqnarray}%
The first equation in eqs.(\ref{9}) can be rewritten as 
\begin{eqnarray}
a &>&1,\alpha \rightarrow 1  \nonumber \\
a &<&1,\alpha \rightarrow 0  \label{10}
\end{eqnarray}%
Note, that $\alpha =1$ corresponds to the matter dominated Universe, and $%
\alpha =0$ to the radiation dominated one.

The dependence of the matter dominated GX-invariant 
\begin{equation}
b_{m}=\frac{\mu _{m}a^{3}G}{c^{2}}-\frac{\pi a}{4}.  \label{D.2}
\end{equation}%
vs
\begin{equation}
\mu_{0}=1.0574\cdot 10^{-28}g\cdot cm^{-3}.
\end{equation}
is shown in Fig.1 ($|b_{m}|$ vs $\mu_{0}$).

\includegraphics{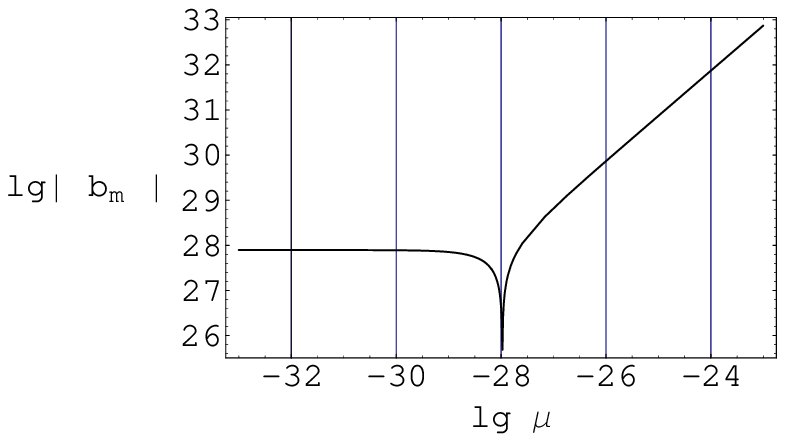}

If $\mu _{0}<10^{-25}g\cdot cm^{-3}$, then $b_{m}<0$. 
So, $b_{m}$ is negative for
today's Universe, while the expansion of the matter dominated phase has
started at $a\approx |b_{m}|$; no matter dominated solutions exist at $%
a<|b_{m}|$. This also means that, the repulsion of vacuum in our Universe
overhelms the attraction of gravity. GX-invariant for radiation $b_{r}$ is 
\begin{equation}
b_{r}=\frac{\mu _{r}a^{4}G}{c^{2}}-\frac{\pi a^{2}}{8},  \label{D.3}
\end{equation}%
and is currently negative, too. However, there is a crucial difference with
the matter dominated case: for radiation the solutions could be
regularizated for $a<|b_{r}|$, while for matter dominated case it is not
allowed.

This indicates that GX-invariants not only explain the symmetries in the
GX-models, but can act as general tools to define the phases of the expansion
of the Universe.


\begin{thebibliography}{9}
\bibitem{GX} V.G. Gurzadyan, S.-S. Xue in: \textquotedblleft From Integrable
Models to Gauge Theories; volume in honor of Sergei
Matinyan\textquotedblright , ed. V.G. Gurzadyan, A.G. Sedrakian, p.177, 
\textit{World Scientific}, 2002;\textit{Mod. Phys. Lett.} \textbf{A18}
(2003) 561 [astro-ph/0105245], see also [astro-ph/0510459]

\bibitem{Zel67} Ya. B. Zeldovich \textit{JETP Lett.} \textbf{6} (1967) 883; 
\textit{Sov. Phys. - Uspekhi} \textbf{95} (1968) 209.

\bibitem{Ver06} G.V. Vereshchagin, \textit{Mod. Phys. Lett.} \textbf{A21}
(2006) 729

\bibitem{Ver06a} G.V. Vereshchagin and G. Yegorian, \textit{Phys. Lett.} 
\textbf{B636} (2006) 150

\bibitem{Ver06b} G.V. Vereshchagin and G. Yegorian, \textit{Class. Quantum
Grav.} \textbf{23} (2006) 5049

\bibitem{Ver06c} G.V. Vereshchagin and G. Yegorian, \textit{astro-ph/}%
0604566.


\end{thebibliography}
\end{document}